\documentclass[onecolumn]{revtex4}
\usepackage{graphicx,epsf}
\usepackage{color}
\usepackage{dcolumn}
\usepackage{amsfonts}
\usepackage{mathrsfs}

\usepackage{subfig}

\begin{document}

\title{On Zero Frequency Zonal Flow and Second Harmonic Generation by Finite Amplitude Energetic Particle Induced Geodesic Acoustic Mode}

\author{Zhiyong Qiu$^{1}$, Ilija Chavdarovski$^2$,   Alessandro Biancalani$^{2}$ and Jintao Cao$^{3}$}

\affiliation{$^1$Institute for    Fusion Theory and Simulation and Department of Physics, Zhejiang University, Hangzhou, P.R.C.\\
$^2$Max-Planck Institute for Plasma Physics, 85748 Garching, Germany\\
$^3$Institute of Physics, Chinese Academy of Science, Beijing, P.R.C.}

\begin{abstract}
Nonlinear self-interaction of finite amplitude energetic particle induced geodesic acoustic mode (EGAM) is investigated using nonlinear gyrokinetic theory. It is found that both zero frequency zonal flow(ZFZF) and second harmonic can be driven by finite amplitude EGAM, with energetic particles (EPs) playing a dominant role in the   nonlinear couplings  through finite orbit width effects. For ZFZF, the effects of EPs on EGAM nonlinear  self-coupling dominate that of the thermal plasmas which are also  present; while the second harmonic  generation is only possible via finite amplitude coupling though EPs.    Our findings may improve the understanding of stabilizing zonal modes, and consequently, drift wave turbulence.
\end{abstract}

\maketitle

\section{Introduction}
Zonal flows (ZFs) \cite{MRosenbluthPRL1998,PDiamondPPCF2005}, or more generally, zonal structures   are toroidally and poloidally symmetric radial corrugations in toroidal devices such as tokamaks. There are two categories of ZFs, i.e., zero frequency zonal flow (ZFZF) \cite{MRosenbluthPRL1998} and geodesic acoustic modes (GAMs)\cite{NWinsorPoF1968,FZoncaEPL2008} peculiar to toroidal plasmas.  ZFZFs are characterized by   zero real frequency and symmetric scalar potentials  determined by trapped ion induced inertia  enhancement \cite{MRosenbluthPRL1998}. On the other hand, GAMs are induced by the thermal plasma compression due to toroidal geometry, with a frequency of the order of sound wave frequency. GAMs are characterized by the up-down anti-symmetric density perturbation, in addition to the predominantly $m=0/n=0$ scaler potential.

ZFs have been extensively studied for two decades due to their potential role  in regulating   turbulence  and the associated anomalous transport \cite{MRosenbluthPRL1998,ZLinScience1998,LChenPoP2000}.  ZFs can be driven unstable by drift waves \cite{LChenPoP2000,ZQiuPoP2014} including drift Alfv\'en waves (DAWs) \cite{LChenPRL2012,ZQiuPoP2016}, and can in turn scatter DWs/DAWs into stable short radial wavelength domain.
It is generally believed that, ZFZFs can more effectively regulate DWs than GAM, due to their low frequency \cite{TSHahmPoP1999}.
While the nonlinear dynamics of DW turbulence depends on the branch ratio between ZFZF and GAM generation, gyrokinetic theory predicts that the cross-section for ZFZF and GAM excitation by DWs are comparable. Thus, the nonlinear DW dynamics may depend on the threshold condition for ZFZF and/or GAM generation, , which itself depends on various plasma parameters such as safety factor and collisionality.

Besides the indirect relation between ZFZF and GAM mediated by DWs, which is sometimes termed as  ``two-predator one-prey" process,  what is adding more complexity to the problem is that, as observed in numerical simulations \cite{HZhangNF2009} and then interpreted by nonlinear gyrokinetic theory \cite{LChenEPL2014},   finite amplitude GAM can directly generate ZFZF, while ZFZF has no feed back on GAM. Hence, one needs to be very careful in interpreting experimental observations of ZFZF. Besides the ZFZF generation, the second harmonic of GAM has alsobeen observed in experiments \cite{YNagashimaPPCF2007,RNAzikianprivatecommunication2009}, while both nonlinear gyrokinetic theory and simulation \cite{HZhangNF2009} and fluid theory \cite{GFuJPP2011} show that second harmonic generation is prohibited due to the cancelation of parallel and perpendicular nonlinearity to the leading order. Thus, the origin for the observed  second harmonic is still open for discussion. One possible interpretation is that energetic particles (EPs) are playing a dominant role in GAM second harmonic generation \cite{GFuJPP2011}, as we will further discuss here.

ZFs are usually  linearly stable due to their symmetric mode structure, and hence they can not be driven unstable by   expansion free energy. However, due to its finite frequency, GAM can be resonantly excited by EPs with the free energy coming from the velocity space anisotropy \cite{RNazikianPRL2008,GFuPRL2008,ZQiuPPCF2010,HWangPRL2013}. Althoug the effect of this EP-induced GAM (EGAM) on DWs are still under investigation \cite{DZarzosoPRL2013,RDumontPPCF2013}, it is proposed as one active control   for DW turbulences.  In this work  we investigate the nonlinear dynamics of finite amplitude EGAM, including both ZFZF and second harmonic generation due to the self couplings of EGAM. Our theory indicates that, both second harmonic and ZFZF can be driven by EGAM, with the finite orbit width (FOW) effects playing a dominant role in the nonlinear couplings. The contribution of resonant EPs to the cross-section of the nonlinear couplings  dominates that of the thermal plasmas.  Our work, thus, may contribute to the understanding of the complex nonlinear dynamics of DWs in the presence of EPs.

The rest of the paper is organized as follows. In Sec. \ref{sec:model}, the theoretical model is given, which is then applied to investigate the ZFZF generation by EGAM in Sec. \ref{sec:ZFZF} and EGAM second harmonic in Sec. \ref{sec:2ndharmonic}. Finally, a brief summary and discussion is given in Sec. \ref{sec:summary}

\section{Theoretical model}
\label{sec:model}

The nonlinear interactions between EGAMs are investigated using nonlinear gyrokinetic theory. For the simplicity of discussion while without loss of generality, we assume $T_e/T_i\ll1$ so that the contribution of $m\neq0$ poloidal sidebands of EGAM, EGAM second harmonic and ZFZF scale potential are negligible \cite{ZQiuPPCF2010,LChenEPL2014}. In this work, for the simplicity of notation, we use $\hat{\Phi}$ for the $m=0$ component of the scalar potential $\delta\phi$. The nonlinear equations describing the nonlinear generation of ZFZF and/or   second harmonic by EGAM can then be derived from the charge quasi-neutrality condition:
\begin{equation}
\sum_{s=i,h}\left\langle\frac{e}{m}\frac{\partial F_{0}}{\partial
E}\Phi+J_k\delta H\right\rangle_s=0,
\label{eq:QN}
\end{equation}
with $\delta H$ being the nonadiabatic part of the perturbed particle response, and can be derived from the  nonlinear gyrokinetic equation \cite{EFriemanPoF1982}
\begin{eqnarray}
\left(\partial_t+\omega_{tr}\partial_{\theta}+ik_rv_{dr}\right)_k\delta H_k=i\omega_k\frac{q_s}{m}J_k\hat{\Phi}\partial_{E}F_0
-\sum_{\mathbf{k}} \delta\mathbf{u}_{k'}\cdot\nabla\delta H_{k''}-\sum_k\delta \dot{E}_{k'}\partial_E\delta f_{k''}.\label{eq:NLGKE}
\end{eqnarray}

A large aspect-ratio axisymmetric tokamak is assumed here, with   the
equilibrium magnetic field given by
$\mathbf{B}_0=B_0(\mathbf{e}_{\xi}/(1+\epsilon\cos\theta)+(\epsilon/q)\mathbf{e}_{\theta})$, where $\xi$ and $\theta$ are, respectively,the toroidal and poloidal
angles of the torus,  $\epsilon=r/R_0\ll1$ is the inverse aspect
ratio, $r$ and $R_0$ are, respectively, the minor and major radii and  $(r,\theta,\xi)$ are straight-field-line toroidal flux coordinates.  Meanwhile,
$v_{dr}=(v^2_{\perp}/2+v^2_{\parallel})/(\Omega R_0)\sin\theta\equiv\hat{v}_{dr}\sin\theta$ is the magnetic drift velocity associated with the geodesic curvature, $\omega_{tr}\equiv v_{\parallel}/(qR_0)$ is the transit frequency, $\sum_k\equiv\sum_{\mathbf{k}=\mathbf{k'}+\mathbf{k''}}$, $\delta \mathbf{u}=\mathbf{b}\times\nabla J_k\hat{\Phi}_k/\Omega$ is the electric field drift velocity,  $\Omega=q_sB/mc$ is the
gyrofrequency,   $J_k\equiv J_0(k_{\perp}\rho_L)$ is the Bessel
function accounting for   finite Larmor radius (FLR) effects, $k_{\perp}$ is the
perpendicular wave vector,
$\rho_L=mcv_{\perp}/q_sB$ is the Larmor radius,
$E=(v^2_{\parallel}+v^2_{\perp})/2$,   and $\delta \dot{E}=q_sv_{dr}\partial_{r}\hat{\Phi}$ corresponds to particle energy change due to magnetic drift in radial direction. The three terms on the right hand side of equation (\ref{eq:NLGKE}) correspond to, respectively,  free energy in phase space, perpendicular nonlinearity and parallel nonlinearity.
Note that, though written explicitly in equation (\ref{eq:NLGKE}), the parallel nonlinearity will not be kept in our derivations  since  it corresponds to a long time scale (slow) process, and   nonlinearity on this time scale  will be neglected systematically.

The linear EP response to EGAM can be derived by transforming equation (\ref{eq:NLGKE}) into the drift orbit center coordinate. Assuming large aspect ratio tokamak and well circulating EPs, and taking $\delta H_h=e^{i\Lambda}\delta H_{dh}$ with $\Lambda\equiv\hat{\Lambda}\cos\theta$ satisfying
$\omega_{tr}\partial_{\theta}\Lambda+\omega_d\sin\theta=0$, we then have
\begin{eqnarray}
\left(\partial_t+\omega_{tr}\partial_{\theta}\right)\delta H_{dh}=-(e/m)\partial_E F_{0h}J_Ge^{-i\Lambda}\partial_t\hat{\Phi}_G,
\end{eqnarray}
and the linear EP response to GAM is then
\begin{eqnarray}
\delta H_{dh}=-\frac{e}{m}\partial_EF_{0h}J_G \sum_l\frac{\omega}{\omega-l\omega_{tr}}(-i)^lJ_l(\hat{\Lambda})e^{il\theta}\hat{\Phi}_G.\label{eq:EP_d_EGAM}
\end{eqnarray}
Note again that in deriving equation (\ref{eq:EP_d_EGAM}),   the $\exp{(iz\cos\theta)}=\sum_{l=-\infty}^{\infty}i^lJ_l(z)\exp(il\theta)$ expansion was used.  Here, $e^{i\Lambda}$ is the operator for coordinate transformation from EP drift orbit center to guiding center,  $\hat{\Lambda}=\hat{\omega_{d}}/\omega_{tr}=k_r\hat{\rho}_d$, and $\hat{\rho}_d=\hat{v}_d/\omega_{tr}$.
The EP response to GAM is then
\begin{eqnarray}
\delta H_{h}=-\frac{e}{m}\partial_EF_{0h}J_G \sum_pi^pJ_p(\hat{\Lambda})e^{ip\theta}\sum_l\frac{\omega}{\omega-l\omega_{tr}}(-i)^lJ_l(\hat{\Lambda})e^{il\theta}\hat{\Phi}_G.\label{eq:EP_EGAM}
\end{eqnarray}
The dispersion relation of EGAM can then be derived by substituting  equation (\ref{eq:EP_EGAM}) into the quasi neutrality condition \cite{ZQiuPPCF2010}, while linear thermal plasma responses to GAM are derived in \cite{ZQiuPPCF2009} for parameter regime relevant to realistic tokamak experiments.

It is clear from last equation that, the linear drive of EGAM comes from the harmonics of transit resonances $\omega=l\omega_{tr}$, with the ``number" of resonant EPs proportional to $J^2_l(\hat{\Lambda})$. As a result, in the small drift orbit limit with $|\hat{\Lambda}|\ll1$, the $l=\pm1$ transit resonances dominate; while for relatively big drift orbits due to short wavelength, higher EP energy,  and/or large safety factor $q$, higher order resonances may also play an important role \cite{HSugamaJPP2006,ZQiuPPCF2009,XXuPRL2008}. It is also evident from wave-particle resonance condition that the optimal ordering of EGAM drive is $T_h/T_i\sim q^2$ \cite{ZQiuPPCF2010}.

The linear particle responses to ZFZF is given in detail in Ref. \cite{LChenEPL2014}. Nonlinear thermal ion and electron responses to  ZFZF and GAM second harmonic are derived in, respectively, Refs. \cite{LChenEPL2014} and \cite{HZhangNF2009}. In this work, we will present the detailed derivation of nonlinear EP response to both ZFZF and/or EGAM second harmonic;harmonic, while we refer the readers to Refs. \cite{LChenEPL2014} and \cite{HZhangNF2009} for the analysis of thermal plasma response.

\section{ZFZF generation by EGAM}
\label{sec:ZFZF}

The nonlinear thermal plasma contribution to ZFZF is derived in Ref. \cite{LChenEPL2014}, and we will focus here on the nonlinear EP contribution. Similarly to thermal plamsa, the nonlinear EP response to ZFZF can be derived by transforming into drift orbit center coordinate. Taking $\delta H^{NL}_{Z,h}=e^{i\Lambda_Z}\delta H^{NL}_{dZ,h}$, and keeping only the dominant perpendicular nonlinearity,  we have
\begin{eqnarray}
(\partial_t+\omega_{tr}\partial_{\theta})\delta H^{NL}_{dZ,h}=-e^{-i\Lambda_Z}\sum_Z\delta \mathbf{u}_E\cdot\nabla\delta H.
\end{eqnarray}

Noting that $|\widetilde{\delta H^{NL}_{dZ,h}}/\overline{\delta H^{NL}_{dZ,h}}|\simeq |\omega_Z/\omega_{tr,h}|\ll1$ with $\widetilde{(\cdots)}$ and $\overline{(\cdots)}$ denoting  respectively, $m\neq0$ and $m=0$ poloidal harmonics, we then have $\delta H^{NL}_{dZ,h}\simeq\overline{\delta H^{NL}_{dZ,h}}$, and
\begin{eqnarray}
\partial_t\overline{\delta H^{NL}_{dZ,h}}&=&-\overline{e^{-ik_Z \rho_d}\sum_Z\delta \mathbf{u}_E\cdot\nabla\delta H}\nonumber\\
&=&-\frac{e}{m}\frac{c}{B_0}\partial_E F_{0h}\overline{\left(1-ik_Z\rho_d\right)\sum_Z \frac{\delta E_r}{r}\partial_{\theta}\sum_p\sum_l \frac{\omega}{\omega-l\omega_{tr}}i^{p-l}J_p(\hat{\Lambda})J_l(\hat{\Lambda})e^{i(p+l)\theta}\hat{\Phi}_G}  \label{eq:EP_ZFZF_drift}.
\end{eqnarray}

In equation (\ref{eq:EP_ZFZF_drift}), the $J_G$-s denoting FLR effects are systematically neglected in comparison to FOW effects.
Noting that resonant EPs dominate the nonlinear coupling \cite{ZQiuPoP2016}, and assuming $|k_r\rho_{d,h}|\ll1$ for EGAMs typically with global mode structure,  the dominant contribution comes from $p=0$ and $l=\pm1$ harmonics. One then have, after tedious but straightforward algebra,
\begin{eqnarray}
\partial_t\overline{\delta H^{NL}_{dZ,h}}&=&-2i\frac{e}{m}\frac{c}{B_0}\partial_E F_{0h}\overline{\rho_d\hat{v}_d\cos\theta}J_0(\hat{\Lambda})\frac{\partial}{\partial r}\left(\frac{|\delta E_r|^2}{r}\right)\left(\frac{\omega_0}{\omega^2_0-\omega^2_{tr}}-\frac{\omega^*_0}{(\omega^*_0)^2-\omega^2_{tr}}\right).
\end{eqnarray}
Here,  $\omega_0=\omega_{0r}+i\gamma$, $\omega^*_0=\omega_{0r}-i\gamma$,  and hence the term in the bracket can then be reduced to
\begin{eqnarray}
\left(\frac{\omega_0}{\omega^2_0-\omega^2_{tr}}-\frac{\omega^*_0}{(\omega^*_0)^2-\omega^2_{tr}}\right)\simeq-i\pi\delta (\omega_{0r}-\omega_{tr}).\label{eq:res_EP_ZFZF}
\end{eqnarray}
In deriving equation (\ref{eq:res_EP_ZFZF}), only resonant EP contributions were kept. Noting again $\delta H^{NL}_{Z,h}=e^{i\Lambda_Z} \delta H^{NL}_{dZ,h}$ and $\delta H^{NL}_{dZ,h}\simeq\overline{\delta H^{NL}_{dZ,h}}$, we then have
\begin{eqnarray}
\partial_t\overline{\delta H^{NL}_{Z,h}}\simeq-2\pi\frac{e}{m}\frac{c}{B_0}\partial_EF_{0h}J^2_0(\hat{\Lambda})\overline{\rho_d\hat{v}_d\cos\theta}\delta (\omega_{0r}-\omega_{tr})\frac{\partial}{\partial r}\left(\frac{|\delta E_r|^2}{r}\right).
\end{eqnarray}
In the previous equation,
the resonant EPs play dominant role, while EGAM is linearly growing, so that  $|\delta E_r|^2\propto \exp(2\gamma_Lt)$ with $\gamma_L$ being the EGAM linear growth rate. We then have  $\partial_t\overline{\delta H^{NL}_{Z,h}}=2\gamma_L\overline{\delta H^{NL}_{Z,h}}$, and substituting $\delta H^{NL}_{Z,h}$ into the quasi-neutrality condition,   we obtain
\begin{eqnarray}
\chi_{iZ}\delta\phi_Z=-\pi\frac{T_i}{n_0m_i}\frac{c}{B_0}\frac{1}{\gamma_L}\left\langle \partial_EF_{0h}\overline{\rho_d\hat{v}_d\cos\theta}J^2_0(\hat{\Lambda})\delta (\omega_{0r}-\omega_{tr})\right\rangle \frac{\partial}{\partial r}\left(\frac{|\delta E_r|^2}{r}\right).\label{eq:ZFZF}
\end{eqnarray}
Here, $\chi_{iZ}\simeq 1.6 k^2_Z\rho^2_iq^2/\sqrt{\epsilon}$ is the neoclassical polarization of ZFZF:
\begin{eqnarray}
\chi_{iZ}\overline{\delta\phi_Z}\equiv\left(1-\left\langle \frac{F_0}{n_i}J^2_Z\left|\overline{e^{ik_Z\rho_d}}\right|^2\right\rangle\right)\overline{\delta\phi_Z},\nonumber
\end{eqnarray}
with the dominant contribution coming from trapped thermal ions \cite{MRosenbluthPRL1998}. In equation (\ref{eq:ZFZF}), the   thermal ions contribution to ZFZF generation,  derived in Ref. \cite{LChenEPL2014}, is of order $O((T_h/T_i)^2(\gamma_L/\omega_G)^2)$ smaller    compared to the contribution of resonant EPs, and is therefore neglected here.

\section{Second harmonic generation by EGAM}
\label{sec:2ndharmonic}

Second harmonic of GAM was observed in JFT-2M \cite{YNagashimaPPCF2007}, Diii-D \cite{RNAzikianprivatecommunication2009} and more recently, Asdex Upgrade \cite{LHorvathNF2016}  experiments, and has been investigated using both gyrokinetic theory and simulation \cite{HZhangNF2009}
and also by fluid theory \cite{GFuJPP2011}. When parallel nonlinearity, which is typically   much smaller than perpendicular nonlinearity, is turned off, finite   GAM second harmonic generation is observed in GTC simulation, and the observed   polarization and mode amplitude agree quantitatively with the nonlinear gyrokinetic theory \cite{HZhangNF2009}; however, when parallel nonlinearity is turned on, the GAM second harmonic generation is reduced by one order \cite{HZhangNF2009}. Theory based on phase-space-conserved form of  gyrokinetic equation \cite{ABrizardPoP1995,TSHahmPoF1989} shows that, the parallel nonlinearity cancels exactly the perpendicular nonlinearity, which for the case of GAM self coupling,  contributes only through weak toroidal coupling \cite{HZhangNF2009}. The null-generation of second harmonic electric field by GAM has been   confirmed in fluid theory \cite{GFuJPP2011}, which also shows presence of  a finite second harmonic density. It  can  be then conjectured, using the analogy of EGAM to the well-known beam-plasma instability, that EPs  play a dominant role in  the second harmonic generation \cite{ZQiuPST2011}, and a theory based on fluid-drift kinetic hybrid model confirms that EPs indeed play an important role in second harmonic generation \cite{GFuJPP2011}, where only resonant EPs are considered under small orbit expansion to focus on the effects on resonant EPs.

In this section, we will re-visit the second harmonic generation by EGAM, with the generalized expression of EP contribution to second harmonic properly treated.  Again, we will focus on the derivation of  nonlinear EPs response to EGAM second harmonic, since as is shown in Ref. \cite{HZhangNF2009} that thermal plasma contribution cancels exactly in the lowest order.
The   EP response to EGAM second harmonic can be derived from the nonlinear gyrokinetic equation:
\begin{eqnarray}
\left(\partial_t+\omega_{tr}\partial_{\theta}+ik_{r,S}\hat{v}_{d}\sin\theta\right)\delta H^{NL}_{S,h}=-\sum_S\delta \mathbf{u}_{E,k'}\cdot\nabla\delta H_{k''},
\end{eqnarray}
with the subscript ``S" denoting second harmonic,  $\sum_S\equiv\sum_{\mathbf{k_S}=\mathbf{k'}+\mathbf{k''}}$, and here $k'$ and $k''$ are both EGAMs. Again, taking $\delta H^{NL}_{S,h}=\exp(i\Lambda_S)\delta H^{NL}_{dS}$, with $\Lambda_S\equiv k_{r,S}(\hat{v}_d/\omega_{tr})\cos\theta\equiv\hat{\Lambda}_S\cos\theta$, one then has the following equation for EP drift orbit center density:
\begin{eqnarray}
\left(\partial_t+\omega_{tr}\partial_{\theta}\right)\delta H^{NL}_{dS,h}&=&-e^{-i\Lambda_S}\sum_S\delta \mathbf{u}_{E}\cdot\nabla\delta H\nonumber\\
&=&-e^{-i\Lambda_S}\frac{e}{m}\partial_EF_{0h}\sum_S \frac{\delta u_{E,\theta}}{r}\frac{\partial}{\partial\theta}\sum_{p,l} i^{p-l}e^{i(p+l)\theta}J_p(\hat{\Lambda})J_l(\hat{\Lambda})\frac{\omega}{\omega-l\omega_{tr}}\hat{\Phi}_G,\label{eq:13}
\end{eqnarray}
which yields
\begin{eqnarray}
\delta H^{NL}_{dS,h}=\frac{e}{m}\partial_EF_{0h}\sum_{p,\xi,l}\frac{p+l}{\omega_S-(p+\xi+l)\omega_{tr}}i^{p-\xi-l}e^{i(p+\xi+l)\theta}J_{\xi}(\hat{\Lambda}_s)J_l(\hat{\Lambda})J_p(\hat{\Lambda})
\sum_S\frac{\delta u_{E,\theta}}{r}\frac{\omega}{\omega-l\omega_{tr}}\hat{\Phi}_G.
\end{eqnarray}
The general expression of nonlinear EP response to EGAM second harmonic can then be written as:
\begin{eqnarray}
\delta H^{NL}_{S,h}=ik_r\frac{c}{B_0}\partial_EF_{0h}\sum_{\eta,\xi,p,l}\frac{p+l}{\omega_S-(p+\xi+l)\omega_{tr}}i^{\eta+p-\xi-l}e^{i(\eta+p+\xi+l)\theta}J_{\eta}(\hat{\Lambda}_S)J_{\xi}(\hat{\Lambda}_S)J_l(\hat{\Lambda})J_p(\hat{\Lambda})
 \frac{\omega}{\omega-l\omega_{tr}}\frac{\hat{\Phi}_G\hat{\Phi}_G}{r}.\label{eq:2nd_EP_general}
\end{eqnarray}

Substituting equation (\ref{eq:2nd_EP_general}) into the surface averaged quasi-neutrality condition, we obtain the equation for EGAM second harmonic generation:
\begin{eqnarray}
b_S\mathscr{E}_{EGAM}(\omega_S)\frac{en_0}{T_i}\hat{\Phi}_S=-\left\langle\overline{\delta H^{NL}_{s,h}} \right\rangle,\label{eq:2nd_DR}
\end{eqnarray}
where $b_S\equiv k^2_{r,S}\rho^2_{L,h}/2$, and $\mathscr{E}_{EGAM}(\omega_S)$ is the linear dielectric function of EGAM at $\omega=\omega_S$:
\begin{eqnarray}
\mathscr{E}_{EGAM}(\omega_S)=-1+\frac{\omega^2_G}{\omega^2_S}-\frac{T_i}{n_0m_ib_S}\left\langle\frac{\partial F_{0h}}{\partial E}\left(1-\sum_l\frac{J^2_l(\hat{\Lambda}_S)\omega_S}{\omega_S-l\omega_{tr}}\right)\right\rangle.
\end{eqnarray}
Equation (\ref{eq:2nd_DR}) is the linear  dispersion relation of EGAM second harmonic  valid for arbitrary   drift orbit width ($|k_r\hat{\rho}_{d,h}|$), which contains  summation over all the transit harmonics, and hence requires numerical solution.
We note that, the general dispersion relation for EGAM second harmonic generation  derived here,  will recover the result  of Ref. \cite{GFuJPP2011} in the proper limit, i.e., with $|\Lambda|\ll1$ and only contribution of resonant EPs taken into account.
For EGAM typically with a global mode structure, i.e., $|\Lambda|\ll1$, the contribution dominates for small $|\eta|+|\xi|+|p|+|l|$. Also, $l=\pm1$ is taken for the strongest linear EGAM drive, and $p+l\neq0$  is required for non-vanishing nonlinear EP response to EGAM second harmonic (the $\partial/\partial\theta$ operator on the right hand side of equation (\ref{eq:13})). With these selection rules in mind, we then have,
\begin{eqnarray}
\overline{\delta H^{NL}_{S,h}} \simeq -2ik_r\frac{c}{B_0}\partial_EF_{0h} J_0(\hat{\Lambda}_S)J_1(\hat{\Lambda}_S)J_0(\hat{\Lambda})J_1(\hat{\Lambda}) \frac{\hat{\Phi}_G \hat{\Phi}_G }{r} \left(\frac{2\omega^2}{\omega_S(\omega^2-\omega^2_{tr})}-\frac{2\omega(\omega\omega_S+\omega^2_{tr})}{(\omega^2-\omega^2_{tr})(\omega^2_S-\omega^2_{tr})}\right).\label{eq:EP_2nd}
\end{eqnarray}
Equation (\ref{eq:EP_2nd}) can be further simplified, noting $\omega_S=2\omega\simeq2\omega_{tr}$ and $\hat{\Lambda}_S=2\hat{\Lambda}$, and we have
\begin{eqnarray}
\overline{\delta H^{NL}_{S,h}}\simeq i\frac{c}{B_0}k_r\hat{\Lambda}^2\frac{\partial F_{0h}}{\partial E}\frac{\omega}{\omega^2-\omega^2_{tr}}\frac{\hat{\Phi}\hat{\Phi}}{r}.\label{eq:2nd_EP_reduced}
\end{eqnarray}

In deriving equation (\ref{eq:2nd_EP_reduced}), $\omega_S=2\omega\simeq2\omega_{tr}$ was applied to simplify the expression. Substituting equation (\ref{eq:2nd_EP_reduced}) into quasi-neutrality condition for EGAM second harmonic, we then obtain:
\begin{eqnarray}
b_S\hat{\mathscr{E}}_{EGAM}(\omega_S)\hat{\Phi}_S=- \frac{ik_rT_i}{n_0m\Omega}\left\langle \hat{\Lambda}^2\frac{\partial F_{0h}}{\partial E} \frac{\omega}{\omega^2-\omega^2_{tr}}\right\rangle \frac{\hat{\Phi}_G\hat{\Phi}_G}{r}\label{eq:2nd_DR_reduced}
\end{eqnarray}
with $\hat{\mathscr{E}}_{EGAM}(\omega_S)$ being the linear EGAM second harmonic dispersion relation in the small orbit limit, and its expression given by
\begin{eqnarray}
\hat{\mathscr{E}}_{EGAM}(\omega_S)\equiv -1+\frac{\omega^2_G}{\omega^2_S}+\frac{T_i}{n_0m_ib_S}\left\langle\frac{\partial F_{0h}}{\partial E} \sum_{l=\pm1,\pm2}\frac{J^2_l(\hat{\Lambda}_S)\omega_S}{\omega_S-l\omega_{tr}} \right\rangle.\label{eq:2nd_DR_linear_reduced}
\end{eqnarray}
In the expression for $\hat{\mathscr{E}}_{EGAM}(\omega_S)$, the $l=\pm2$ transit resonances are kept, in addition to the $l=\pm1$ transit resonances that dominate linear EGAM excitation,  since for EGAM second harmonic, with $\omega_S=2\omega\simeq 2\omega_{tr}$, the contribution of $l=\pm2$ resonances could be comparable with those of $l=\pm1$, even though  $J^2_2(\hat{\Lambda})\ll J^2_1(\hat{\Lambda})$ in the small orbit limit. Note that in the equation (52) of Ref. \cite{GFuJPP2011}, $\omega_{EGAM}$ is also a function of $\omega_S$ ($\omega_2$ using the notation of Ref. \cite{GFuJPP2011}), as we showed in equation (\ref{eq:2nd_DR_linear_reduced}). Equation (\ref{eq:2nd_DR_reduced}) can then be applied to explain experiments/simulations on EGAM second harmonic generation, by direct substituting parameters into the nonlinear dispersion relation.

\section{Summary and discussion}
\label{sec:summary}

In summary, in this work, we have derived the  dispersion relations describing nonlinear generation of ZFZF and second harmonic by finite amplitude EGAM, and shown that EPs play dominant roles in both processes. For ZFZF, it was shown in Ref. \cite{LChenEPL2014} that, in the absence of EPs, finite amplitude  GAM can drive ZFZF, with the thermal ion FOW effects playing a dominant role. When the EPs are taken into account, it is found  that the effects of the resonant EPs induced by the symmetry breaking become dominant, despite their low density.  On the other hand, it is shown by Refs. \cite{HZhangNF2009} and \cite{GFuJPP2011} that, second harmonic scalar potential cannot be driven by finite amplitude GAM due to the cancelation of parallel and perpendicular nonlinearities. When the effects of resonant EPs are taken into account in the long wavelength limit, Ref. \cite{GFuJPP2011} shows that second harmonic can be driven by finite amplitude EGAM. In this work, the nonlinear response of EPs to EGAM second harmonic is derived systematically for arbitrary wavelength (compared to EP drift orbit) by coordinate transforming into EP drift orbit center, which is then used to derive the nonlinear dispersion relation of EGAM second harmonic. It is shown that, the finite coupling comes from EP finite orbit width effect; instead of toroidicity for the perpendicular nonlinearity of thermal ions.  The generation of ZFZF and second harmonic, has a potential to give  further insight into the nonlinear dynamics of turbulence and hence  the corresponding transport.

This work is supported by the National Magnet Confinement Fusion Research Program under Grants Nos.  2013GB104004  and   2013GB111004,
and the National Science Foundation of China under grant Nos.  11575157.

\end{document}